\documentclass[preprint,showpacs,preprintnumbers,amsmath,amssymb,nofootinbib]{revtex4}

\usepackage{graphicx}
\usepackage{dcolumn}
\usepackage{bm}
\usepackage{epstopdf}

\newcommand{\nn}{\nonumber}

\newcommand{\be}{\begin{eqnarray}}
\newcommand{\ee}{\end{eqnarray}}
\catcode`\@=11
\def\lsim{\mathrel{\mathpalette\@versim<}}
\def\gsim{\mathrel{\mathpalette\@versim>}}
\def\@versim#1#2{\vcenter{\offinterlineskip
\ialign{$\m@th#1\hfil##\hfil$\crcr#2\crcr\sim\crcr } }}
\catcode`\@=12

\usepackage{color}

\begin{document}
\vspace{2cm}
\preprint{KANAZAWA-08-09}

\title{Testing Flavor Symmetries by
B-Factory}

\author{Takeshi Araki$^a$}
\author{Jisuke Kubo$^b$}

\affiliation{
$^a$Department of Physics, National Tsing Hua University,
Hsinchu, Taiwan 300\\
$^b$Institute for Theoretical Physics, Kanazawa
University, Kanazawa 920-1192, Japan
\vspace{3cm}
}

\begin{abstract}
The Cabibbo-Kobayashi-Maskawa (CKM) parameters
are investigated in detail in recent predictive models which are based
on low-energy non-abelian discrete family symmetries.
Some of the models can already be excluded at the present
precision of the determination of the CKM parameters,
while some of them seem to survive.
We find that to make the uncertainties of the theoretical values
comparable with the assumed uncertainties
of $\sim 1^{\circ}$ and $\sim 2^{\circ}$ 
in $\phi_2(\alpha)$ and $\phi_3(\gamma)$, respectively,  at about 50 inverse 
atto barn achieved at a future 
B factory, it is necessary to reduce the uncertainties in the quark masses,
especially that of the strange quark mass by more than 60\%.

\end{abstract}

\pacs{12.60.Jv,11.30.Hv, 12.15.Ff, 14.60.Pq, 02.20.Df }

\maketitle

\section{Introduction}
The  success of the standard model (SM)  suggests
that we are very close to a more fundamental
theory for elementary particle physics.
Yet, we do not know how the SM should be extended,
except that
the mass of neutrinos  with their mixing and 
a dark matter candidate have to be incorporated in the extension.
The Higgs sector of the SM indicates that the
extension may take place around TeV scale, and
supersymmetry is widely believed to
be  the best candidate to increase the natural energy scale
of the theory.
Yet another problem is the Yukawa sector,
because the most of the free parameters of the SM
are involved there and the SM does not provide with a principle 
how to fix its structure.
Moreover, simple supersymmetry does not soften
the problem of the Yukawa sector.

A  natural way to provide with a principle
for the Yukawa sector
is the introduction of a family symmetry.
A family symmetry is not necessarily
  adequate to explain the observed hierarchy
  of the fermion masses. It can however relate the 
  fermion masses and mixing parameters.
  That is,  mixing parameters  may be related to  mass ratios.
Note that  the classic relations such as
$\sin\theta_C\simeq \sqrt{m_d/m_s}$
  \cite{Gatto:1968ss,Cabibbo:1968vn,Fritzsch:1977za}, or
  $|V_{ub}/V_{cb}|\simeq  \sqrt{m_u/m_c}$
   \cite{Hall:1993ni,Roberts:2001zy}
   \cite{Ramond:1993kv} had not been derived  from a family symmetry.
Recently there have been a growing number of interests in family symmetries.
Most of the recent papers deal with the large neutrino
 mixing (see for instance 
 \cite{Altarelli:2007gb,Ma:2004pt,Lam:2008sh}), because
 a large mixing may be associated with a family symmetry.
 As for the quark masses and mixing, 
 tremendous works on their ansatz
have been done (see for instance \cite{Fritzsch:1999ee}).
However,
there is an almost no-go theorem \cite{Koide:2004rd} saying
that there exists no viable low-energy family symmetry 
in the SM to understand the fermion mass matrices.
Therefore, if the fermion mass matrices 
should be derived from a family symmetry, one has to extend the SM.
One of the possibilities is to extend  the Higgs sector
such that it also forms a family \cite{Pakvasa:1977in}.
 In fact a  renewed interest in 
this approach to the quark sector
  has  been recently aroused
  \cite{Ma:2002yp}-\cite{Ishimori:2008fi}.

 In this paper we are interested in predictive flavor models with a low-energy
  discrete family symmetry that are testable at future
  B factories such as SuperKEKB \cite{Akeroyd:2004mj}
  and Super Flavour Factory \cite{Browder:2007gg}.
   We met a set of the following selection criteria for the models:\\
  1) The family symmetry should be a low-energy symmetry.
  That is, we do not include here family symmetries
  at GUT scale\footnote{Recent GUT models with a family symmetry
  can be found in \cite{Ma:2005tr}-\cite{Ishimori:2008fi}.}.\\
  2) The family symmetry should not be hardly
broken. If it is hardly broken, there is in general no
quantitative prediction of the symmetry.
So, we do not consider textures which 
are not supported by a symmetry.\\
3) The model should describe 
10 observables, six quark masses and four 
Cabibbo-Kobayashi-Maskawa (CKM) parameters,
by less than 10 parameters.\\
4) The model should be renormalizable.\\
 To our knowledge  there are only four models 
proposed in \cite{Lavoura:2005kx,Chen:2005jt,Babu:2004tn}
that satisfy the selection criteria above.

In the next section we will start by summarizing
the  present values of the CKM parameters and quark masses.
We will  find that the  uncertainties  in the light quark masses
have to be largely reduced
to make the  uncertainties 
of the theoretical values of the CKM parameters
comparable with the assumed experimental uncertainties
of  future B factories.
The lattice calculation
\cite{Blum:2007cy} is indeed an on-going project 
to reduce  the uncertainties in the light quark masses.
In this paper, however,  we will not use the quark mass values
given in \cite{Blum:2007cy}, because 
the uncertainties due to the absence of the sea strange quarks
have not been included.
   
\section{Quark masses and CKM parameters}
Our strategy to compare
the theoretical values with the experimental ones
is to use the six quark masses along with two 
of the CKM parameters to plot
the theoretical values in the plane defined by other two CKM parameters.
From the quark masses in the  $\overline{\mbox{MS}}$ scheme
given in Particle Data Group 2009 \cite{Amsler:2008zz}
we obtain the quark masses at $\mu=M_Z$ \cite{Kim:2004ki}
\begin{eqnarray}
& & m_u(M_Z)=0.98\sim 2.15\ {\rm MeV}~,~
m_d(M_Z)=2.28\sim 3.90\ {\rm MeV},\nn\\
& & m_s(M_Z)=45.5\sim 84.5\ {\rm MeV}~,~
m_c(M_Z)=0.65\sim 0.75\ {\rm GeV},
\label{quarkmasses}\\
& & m_b(M_Z)=2.85\sim 3.02\ {\rm GeV}~,~
m_t(M_Z)=179.2\sim 183.7\ {\rm GeV}.\nn
\end{eqnarray}
The input parameters we shall use are the ratios at $\mu=M_Z$, i.e.
\begin{eqnarray}
& & m_u / m_t=(0.54\sim 1.18)\times 10^{-5}~,~
m_c / m_t=(0.36\sim 0.41)\times 10^{-2},\nn\\
& & m_d / m_b=(0.79\sim 1.34)\times 10^{-3}~,~
m_s / m_b=(0.16\sim 0.29)\times 10^{-1}.
\label{massratio}
\end{eqnarray}
Given the  ratios above, there are two independent
mass scales, one  for the up quarks and the other
 for the down quarks: if necessary we use the fixed values 
 $m_b(M_Z)=2.9\ {\rm GeV}\ ,\ m_t(M_Z)=181.4\ {\rm GeV}$.
We further require that \cite{Amsler:2008zz}
\begin{eqnarray}
&& m_s / m_d=17\sim 22~,~
\frac{m_u + m_d}{2}=2.5\sim  5.5\ {\rm MeV}~,~
\frac{2 m_s}{m_u + m_d}=25\sim  30~,~\nn\\
& & m_u / m_d=0.35\sim 0.6~,~
\frac{m_s - (m_u + m_d)/2}{m_d - m_u}=30\sim 50
\end{eqnarray}
are satisfied.
We also use 
\begin{eqnarray}
&& |V_{us}|=0.2236\sim 0.2274~,~
|V_{cb}|=0.0401\sim 0.0423
\label{V-input}
\end{eqnarray}
as the two input parameters of the CKM parameters.
Then the theoretical values are compared with 
 two sets of the experimental values:
\begin{flushleft}
{\bf PDG:}\cite{Amsler:2008zz}
\end{flushleft}
\begin{eqnarray}
 &&\bar{\rho}=0.1425 \pm 0.0235~,~
\bar{\eta}=0.348 \pm 0.016,\nn\\
 &&\phi_2(\alpha)=(88.5 \pm 5.5)^{\circ}~,~
\phi_3(\gamma)=(76 \pm 31)^{\circ}~,~
\sin 2\phi_1(\beta)=(0.681 \pm 0.025), 
\label{exp-v-pdg}\\
 &&
|V_{ub}|=(3.93 \pm 0.36)\times 10^{-3}~,~
|V_{td}/V_{ts}|=0.209 \pm 0.007,\nn
\end{eqnarray}
and
\begin{flushleft}
{\bf UTFit:}\cite{utfit}
\end{flushleft}
\begin{eqnarray}
 &&\bar{\rho}=0.154 \pm 0.022~,~
\bar{\eta}=0.342 \pm 0.014, \nn\\
 &&\phi_2(\alpha)=(92.0 \pm 3.4)^{\circ}~,~
\phi_3(\gamma)=(65.6 \pm 3.3)^{\circ}~,~
\sin 2\phi_1(\beta)=0.695 \pm 0.020,
\label{exp-v-utfit}\\
 &&
|V_{ub}|=(3.60 \pm 0.12)\times 10^{-3}~,~
|V_{td}/V_{ts}|=0.209 \pm 0.0075.\nn
\end{eqnarray}

\section{The models and their CKM parameters}
In this section we consider four different flavor
models with three different family symmetries.
Except the model II of \cite{Chen:2005jt}
these models should be supersymmetric to obtain
desired mass matrices for the quarks.
The family symmetry of the models III and 
IV of \cite{Babu:2004tn} extends to the leptonic sector
so that there are nontrivial, testable predictions 
in that sector, too\footnote{See 
\cite{Kajiyama:2005rk,Kifune:2007fj} for an alternative assignment
of the leptons to obtain the maximal mixing of 
the atmospheric neutrinos.}.
The $SU(2)_L$ doublets
of the quarks and Higgs bosons
are denoted by $Q $ and  $ H^{u,d}$, respectively.
Similarly, $SU(2)_L$ singlets 
of the quarks are denoted by
$U^c$ and $D^c$.
In the supersymmetric models they are superfields, and
they are ordinary fields in the non-supersymmetric case.

\subsection{The model  I \cite{Lavoura:2005kx}}
\begin{table}[t]
\begin{center}
\begin{tabular}{|c||c|c|c|c|c|}\hline
 &
 $Q_{1,2}\ U^c_{1,2}\ D^c_{1,2}$ &
 $ H^u_{1,2}$ &
 $ H^d_{1,2}$ &
 $Q_3\ U^c_3\ D^c_3$ &
 $H^u_3\ H^d_3$ \\ \hline
 $S_3$ & {\bf 2} & {\bf 2} & {\bf 2} & {\bf 1} & {\bf 1} \\ \hline
 $Z_2$ & 1 & $-1$ & 1 & $-1$ & 1 \\ \hline
\end{tabular}
\end{center}
\caption{The $S_3$ and $Z_2$ assignment in the model I. 
\cite{Lavoura:2005kx}}
\label{tab:S3-I}
\end{table}
The first model, which is  proposed in \cite{Lavoura:2005kx}, 
 is the supersymmetric  model
with an $S_3\times Z_2$ family symmetry.
The $S_3\times Z_2$ assignment is given in Table \ref{tab:S3-I}, and
the  $S_3 \times Z_2$ invariant cubic superpotential for the Yukawa 
interactions in the quark sector is given by
\begin{eqnarray}
&&W_q = Y^u_a Q_3 U^c_3 H^u_3
                             + Y^u_b (Q_1 H^u_2 + Q_2 H^u_1) U^c_3\nonumber\\
&&\hspace{1cm}+ Y^u_c Q_3 (U^c_1 H^u_2 + U^c_2 H^u_1 ) 
                             + Y^u_e (Q_1 U^c_2 + Q_2 U^c_1 )H^u_3\nonumber\\
&&\hspace{1cm}+ Y^d_a Q_3 D^c_3 H^d_3
                             + Y^d_e (Q_1 D^c_2 + Q_2 D^c_1) H^d_3
          + Y^d_f \left( Q_2 D^c_2 H^d_2 + Q_1 D^c_1 H^d_1 \right)\ ,
\end{eqnarray}
where the subscripts $1,2$ and $3$ stand for the 
two components of the
 $S_3$ doublet and for the $S_3$ singlet,
respectively.
\begin{figure}[htb]
\includegraphics*[width=0.4\textwidth]{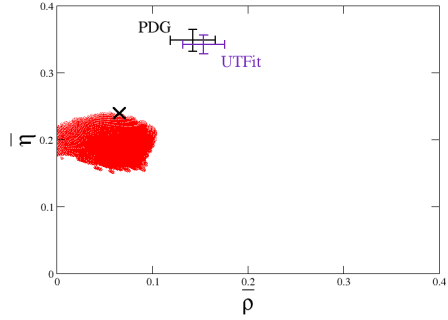}
\caption{\label{rho-eta-s3}\footnotesize
The  theoretical values  of the model I at $90\%$ CL in
the $\bar{\rho}{\bf -}\bar{\eta}$ plane.
 Two experimental values (\ref{exp-v-pdg}) and 
(\ref{exp-v-utfit}) are also plotted.
The best fit point is denoted as $\times$.
 }
\end{figure}
After the spontaneous electroweak symmetry breaking, 
the Higgs doublets should acquire
vacuum expectation values (VEVs) $<H_i^{u,d}>=v_i^{u,d}$.
Then the quark mass matrices are\footnote{
This model is very similar to the ansatz considered in \cite{He:1989eh}.}
\begin{eqnarray}
M_u=
\left(\begin{array}{ccc}
 0 & Y^u_e\ v^u_3 & 0 \\
 Y^u_e\ v^u_3 & 0 & Y^u_b\ v^u_1 \\
 0 & Y^u_c\ v^u_1 & Y^u_a\ v^u_3
\end{array}\right)
\equiv
\left(\begin{array}{ccc}
 0 & x & 0 \\
 x & 0 & z \\
 0 & y & w
\end{array}\right)\ ,
\end{eqnarray}
\begin{eqnarray}
M_d=
\left(\begin{array}{ccc}
 0 & Y^d_e\ v^d_3 & 0 \\
 Y^d_e\ v^d_3 & Y^d_f\ v^d_2 & 0 \\
 0 & 0 & Y^d_a\ v^d_3
\end{array}\right)
\equiv
\left(\begin{array}{ccc}
 0 & p & 0 \\
 p & q & 0 \\
 0 & 0 & m_b
\end{array}\right)\ .
\end{eqnarray}
Note that  $v_2^u = v_1^d =0$ is assumed, and
supersymmetry makes it possible to obtain this relation
naturally \cite{Lavoura:2005kx}.

All the elements of these matrices can be made real and positive
by an appropriate redefinition of the quark fields.
Therefore, the real quark mass matrices can be diagonalized by
orthogonal matrices as follows,
\begin{eqnarray}
&&O_u^T M_u M_u^T O_u = {\rm diag}(m_u^2 , m_c^2 , m_t ^2)\ ,\\
&&O_d^T M_d M_d^T O_d = {\rm diag}(m_d^2 , m_s^2 , m_b ^2)\ ,
\end{eqnarray}
and the CKM matrix is written as
\begin{eqnarray}
V=O_u^T P O_d\ ,
\end{eqnarray}
where $P={\rm diag}(1,e^{i\phi},1)$.
There are only eight independent parameters, so that
we can calculate two remaining physical quantities.

The CKM matrix $V$ can be approximately
obtained in a closed form, and one finds \cite{Lavoura:2005kx}, for instance,
\begin{eqnarray}
 &&\left| \frac{V_{ub}}{V_{cb}}\right| \simeq \frac{xw}{yz}
                                  \simeq\sqrt{\frac{m_u}{m_c}}~,~
         \frac{V_{td}}{V_{ts}}\simeq -\tan\alpha_d 
                      \simeq -\sqrt{\frac{m_d}{m_s}}\ .
\end{eqnarray}
Using the values given
in  (\ref{quarkmasses}), (\ref{massratio}) and
 (\ref{V-input}),  we find
 \begin{eqnarray}
 &&|V_{ub}|\simeq |V_{cb}|\sqrt{\frac{m_u}{m_c}}
           \simeq  (1.4 \sim 2.4)\times 10^{-3} \ ,\\
 &&\left| \frac{V_{td}}{V_{ts}}\right| \simeq \sqrt{\frac{m_d}{m_s}}
           \simeq 0.213 \sim 0.242 \ .
\end{eqnarray}
As we can see from (\ref{exp-v-pdg}) and (\ref{exp-v-utfit}),
the ratio $\left| V_{td}/V_{ts}\right|$ 
is consistent  with the experimental value, while $ |V_{ub}|$ is not.
We have performed a systematic, numerical analysis
of the theoretical values in various planes. 
Fig.~\ref{rho-eta-s3} shows the case of the model I at $90\%$ CL
in
the $\bar{\rho}{\bf -}\bar{\eta}$ plane.
We clearly see that the model I is not consistent with the experimental
observations\footnote{Changing the $Z_2$ assignment appropriately,
one can  interchange the mass matrices for the up and down quarks.
In this case, however, one obtains a negative
$\rho$  \cite{Lavoura:2005kx}, which is excluded experimentally.}.

\subsection{The model  II \cite{Chen:2005jt}}
\begin{table}[t]
\begin{center}
\begin{tabular}{|c||c|c|c|c|c|c|c|}\hline
    & $\bar{Q}_{1,2}\ d_{1,2}$ & $u_{1,2}$ 
    & $H^{d}_{1,2}$ & $H^{u}_{1,2}$
    & $\bar{Q}_{3}\ d_{3}\ u_{3}$
    & $H^{d}_{3}\ H^{u}_{3}$ \\ \hline
  $D_{7}$ & \bf 2 & $\bf 2^{'}$ & \bf 2 & $\bf 2^{''}$ & \bf 1 
              & \bf 1 \\ \hline
\end{tabular}
\end{center}
\caption{$D_7$ assignments of the matter fields.}
\label{tab:ChenMa}
\end{table}

Next we  consider the $D_7$ flavor symmetric model 
of \cite{Chen:2005jt}, where
 a supersymmetric extension is not always necessary in this model.
The $D_7$ assignment is given in Table \ref{tab:ChenMa}, and 
the $D_7$ invariant Yukawa Lagrangian of the quark sector
can be written as
\begin{eqnarray}
{\cal L}_q
&=& Y_a^u (\bar{Q}_{2} H_1^u d_{2} + \bar{Q}_1 H_2^u u_1) 
 + Y_b^u Q_3 H_3^u u_3 \nonumber\\
&& + Y_a^d (\bar{Q}_1 d_2 + \bar{Q}_2 d_1 )H_3^d
 +Y_b^d  (\bar{Q}_1 H_2 + \bar{Q}_2 H_1^d) d_3\nonumber\\
&&+Y_c^d \bar{Q}_3 (H_1^d d_2 + H_2^d d_1)
 +Y_e^d \bar{Q}_3 H_3^d d_3+h.c.\ ,
\end{eqnarray}
which  yields the quark mass matrices of the form
\begin{eqnarray}
&&M_u=
\left( \begin{array}{ccc}
 Y_a^u\ v^u_2 & 0 &0 \\
 0 & Y_a^u\ v^u_1 & 0 \\
 0 & 0 & Y_b^u\ v^u_3
\end{array}\right)
\equiv
\left( \begin{array}{ccc}
 m_u & 0 &0 \\
 0 & m_c & 0 \\
 0 & 0 & m_t
\end{array}\right)\ ,\\
&&M_d=
\left( \begin{array}{ccc}
 0 & Y_a^d\ v^d_3 & Y_b^d\ v^d_2 \\
 Y_a^d\ v^d_3 & 0 & Y_b^d\ v^d_1 \\
 Y_c^d\ v^d_2 & Y_c^d\ v^d_1 & Y_e^d\ v^d_3
\end{array}\right)
\equiv
\left( \begin{array}{ccc}
 0 & a & \xi b \\
 a & 0 & b \\
 \xi c & c & d
\end{array}\right)\ ,
\end{eqnarray}
where $<H^{u,d}_i> \equiv v^{u,d}_i$.
 Except for $\xi$, the parameters $a, b, c$ and $d$ of $M_d$
 can be made real and positive.
That is, there are nine independent parameters in the quark sector, and
we can calculate one physical quantity.
\begin{figure}[htb]
\includegraphics*[width=0.4\textwidth]{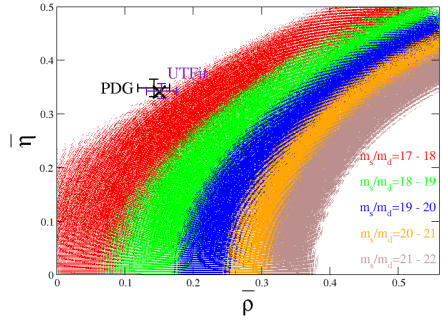}
\caption{\label{rho-eta-d7}\footnotesize
The  theoretical values of the model II at $90\%$ CL in the 
$\bar{\rho}{\bf -}\bar{\eta}$ plane.
Different colors indicate different intervals of the mass ratio
$m_s/m_d$: $17\sim18$ (red),
$18\sim19$ (green), $19\sim20$ (blue), $20\sim21$ (orange), and
$21\sim 22$ (brown).
 Two experimental values (\ref{exp-v-pdg}) and 
(\ref{exp-v-utfit}) are also plotted.
The best fit point of the model II is denoted as $\times$.
}
\end{figure}
The mass matrix of the  down quarks can be diagonalized 
by a unitary matrix
$V_d$ as
\begin{eqnarray}
V_d^{\dag}M_d M_d^{\dag}V_d = {\rm diag}(m_d^2 , m_s^2 , m_b^2)\ ,
\end{eqnarray}
and the CKM matrix is then given by
\begin{eqnarray}
V=V_d\ ,
\end{eqnarray}
because the  mass matrix of  the up quarks $M_u $ is  diagonal,
which is a consequence of the family symmetry.
If one assumes that $a^2 \ll b^2$ and $|\xi|^2 \ll 1$,
one finds various approximate relations such as \cite{Chen:2005jt}:
\begin{eqnarray}
 && m_b \simeq \sqrt{c^2 + d^2}~,~
 m_s \simeq \frac{bc}{\sqrt{c^2 + d^2}}~,~
m_d \simeq \left| \frac{a^2 d}{bc}-2\xi a \right| \ ,\\
 && V_{cb} \simeq \frac{bd}{c^2 + d^2}~,~
V_{us} \simeq -\frac{ad}{bc}+\xi ~,~
V_{ub} \simeq \frac{ac+\xi bd}{c^2 +d^2}\ .
\end{eqnarray}
Instead of using these approximate analytic
expressions, we have performed
systematic numerical analyses.
The theoretical values in the $\bar{\rho}{\bf -}\bar{\eta}$ plane
are plotted in 
Fig. \ref{rho-eta-d7}.
Different colors mean different intervals of the mass ratio
$m_s / m_d$: $17\sim18$ (red),
$18\sim19$ (green), $19\sim20$ (blue), $20\sim21$ (orange), and
$21\sim 22$ (brown). 
The best fit for the set of parameters,
$\sin2\phi_1(\beta),\phi_3(\gamma),
m_d/m_b,m_s/m_b,|V_{us}|,|V_{cb}|$ is given by
\be
  \sin2\phi_1(\beta) &=& 0.682,~\phi_3(\gamma)=62.314^\circ ,~ 
    |V_{us}|=0.2256,~|V_{cb}|=0.0411,\nonumber\\
  m_d/m_b &=&1.120\times10^{-3},~
  m_s/m_b=0.191\times10^{-1}
  \ee
with  $\chi^2/\mbox{dof}  = 0.51$.
As we can see from the figure, the $D_7$ model
is consistent with the experimental observations.
More precise measurements of the CKM parameters
as well as  more precise determinations of 
$m_s/m_d$ are needed to confirm or exclude the model.
If the mass ratio $m_s/m_d$ turns out to be  larger than $18$,
the model may run into  problems.

\section{The model III  \cite {Babu:2004tn} and IV}
\begin{table}[t]
\begin{center}
\begin{tabular}{|c||c|c|c|c|c|c|}\hline
          & $Q_{1,2}$ 
          & $U^{c}_{1,2}\ D^{c}_{1,2}$
          & $H^{u}_{1,2}\ H^{d}_{1,2}$
          & $Q_{3}$
          & $U^{c}_{3}\ D^{c}_{3}$
          & $H^{u}_{3}\ H^{d}_{3}$\\ \hline
 $Q_{6}$  & ${\bf 2}$
          & ${\bf 2}^{'}$
          & ${\bf 2}^{'}$
          & ${\bf 1}^{'}$
          & ${\bf 1}^{'''}$
          & ${\bf 1}^{'''}$\\ \hline
\end{tabular}
\end{center}
\caption{The $Q_6$ assignment of the matter multiplets.
The  assignment is the same
 for the model III \label{tab:BabuKubo} and IV.}

\end{table}
The last two supersymmetric models are based on a $Q_6$ family
symmetry  \cite{Babu:2004tn}.
In contrast to the previous two models, 
the family symmetry of these models extends
to the lepton sector.
The $Q_6$ assignment of  the leptons
given in \cite{Kajiyama:2005rk} indeed leads to the maximal mixing
of the atmospheric neutrinos. Note, however, the parameter space
of the model IV has not been discussed previously.
The finite group $Q_6$ allows complex representations, and
the $Q_6$ assignment of the matter multiplets is given
in Table \ref{tab:BabuKubo}.

The superpotential for  the Yukawa interactions in
the quark sector is given by
\begin{eqnarray}
W_q
&=&Y_a^u Q_3 H_3^u U_3^c + Y_b^u (Q_1 H_2^u+Q_2 H_1^u) U_3^c
+ Y_c^u Q_3 (H_1^u U_2^c-H_2^u U_1^c)\nonumber\\
&&+ Y_e^u (Q_1 U_2^c+Q_2 U_1^c)H_3^u\nonumber\\
&&+Y_a^d Q_3 H_3^d D_3^c + Y_b^d (Q_1 H_2^d+Q_2 H_1^d) D_3^c
+ Y_c^d Q_3 (H_1^d D_2^c-H_2^d D_1^c)\nonumber\\
&&+ Y_e^d (Q_1 D_2^c+Q_2 D_1^c)H_3^d\ .
\end{eqnarray}
To make the model predictive there are two crucial requirements:
(1) the VEV alignment
\begin{eqnarray}
&&<H^{u,d}_1>=<H^{u,d}_2> \equiv v^{u,d}_1~,~
<H^{u,d}_3> \equiv v^{u,d}_3 ,
\end{eqnarray}
which can be achieved by an accidental 
permutation symmetry
$H^{u,d}_1 \leftrightarrow H^{u,d}_2$
 in the Higgs sector, 
and (2) CP is  spontaneously broken.
The second requirement can be relaxed to that
 the Yukawa couplings are real
without contradicting renormalizability\footnote{It has been found 
\cite{Kifune:2007fj} that
  to trigger complex VEVs with the minimal
  content of the chiral supermultiplets given
  in Table \ref{tab:BabuKubo},
  the family symmetry and 
  CP should be  at
  least softly broken.
  The most economic breaking can be achieved by the b-terms
   in the soft-supersymmetry
  breaking sector.}.
 Then the quark mass matrices can be written as
\begin{eqnarray}
&&M_u=
\left( \begin{array}{ccc}
 0 & Y_e^u\ v_3^u & Y_b^u\ v_1^u \\
 Y_e^u\ v_3^u & 0 & Y_b^u\ v_1^u \\
 -Y_c^u\ v_1^u & Y_c^u\ v_1^u & Y_a^u\ v_3^u
\end{array}\right)~,~
M_d=
\left( \begin{array}{ccc}
 0 & Y_e^d\ v_3^d & Y_b^d\ v_1^d \\
 Y_e^d\ v_3^d & 0 & Y_b^d\ v_1^d \\
 -Y_c^d\ v_1^d & Y_c^d\ v_1^d & Y_a^d\ v_3^d
\end{array}\right)\ 
\end{eqnarray}
with complex VEVs.
\begin{figure}[htb]
\includegraphics*[width=0.4\textwidth]{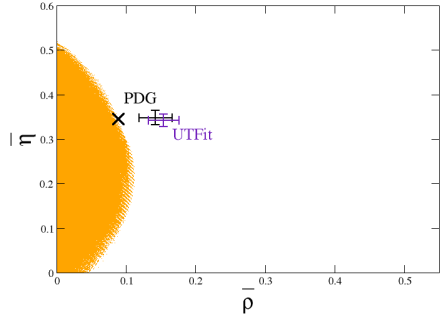}
\caption{\label{s2b-gamma-III}
\footnotesize
The theoretical values of the model III at $90 \%$ CL in the 
 $\bar{\rho}{\bf -}\bar{\eta}$ plane .
  Two experimental values (\ref{exp-v-pdg}) and 
(\ref{exp-v-utfit}) are also plotted.
The best fit point is denoted as $\times$.
 }
\end{figure}
By making an overall $45^{\circ}$ rotation of the $Q_6$ doublets 
$Q, U^c$
and $D^c$ in the space of the family group, 
we obtain nearest neighbor interaction (NNI) type
mass matrices:
\begin{eqnarray}
&&M_u=
\left( \begin{array}{ccc}
 0 & c_u & 0 \\
 -c_u & 0 & b_u\\
 0 & d_u & e_u
\end{array}\right)~,~
M_d=
\left( \begin{array}{ccc}
 0 & c_d & 0 \\
 -c_d & 0 & b_d \\
 0 & d_d & e_d
\end{array}\right)\ .
\label{mumd}
\end{eqnarray}
All the elements of these matrices can be made real by a suitable redefinition
of the quark fields.
Then the real  matrices can be diagonalized by  orthogonal
matrices as 
\begin{eqnarray}
&&O_u^T M_u M_u^T O_u = {\rm diag}(m_u^2 , m_c^2 , m_t ^2),\\
&&O_d^T M_d M_d^T O_d = {\rm diag}(m_d^2 , m_s^2 , m_b ^2),
\end{eqnarray}
and the CKM matrix takes the form
\begin{eqnarray}
V=O_u^T P O_d, 
\end{eqnarray}
where $P={\rm diag}(1,e^{2i\theta},e^{i\theta})$.
The phase rotation matrix $P$ has only  one
angle $\theta$, which is the consequence of a
spontaneously and  softly broken CP.

The full nine dimensional space
of the parameters can be divided into 
two non-equivalent regions.
The difference between  the two regions, the region III 
and IV, can be reduced to the sign of $c_d$,
so that  without loss of generality the other parameters, i.e. $b_u, c_u\dots,
e_d$ in (\ref{mumd}), can be assumed to be  positive real numbers.
(The parameter space for the model IV has not been considered in 
\cite{Babu:2004tn}.)
Accordingly, we define two models; the model III 
for positive $c_d$ and IV for negative $c_d$.
\begin{figure}[htb]
\includegraphics*[width=0.4\textwidth]{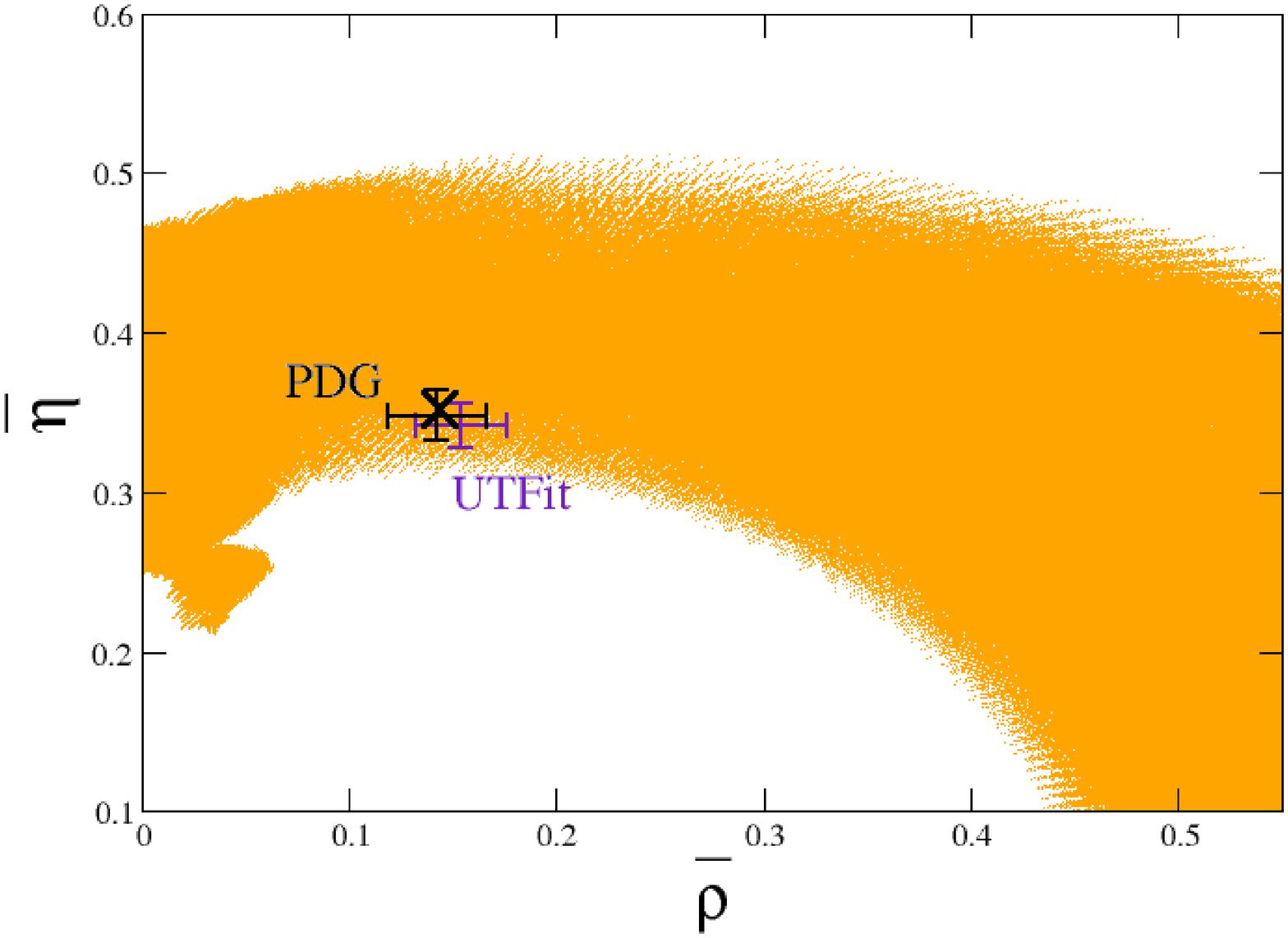}
\includegraphics*[width=0.4\textwidth]{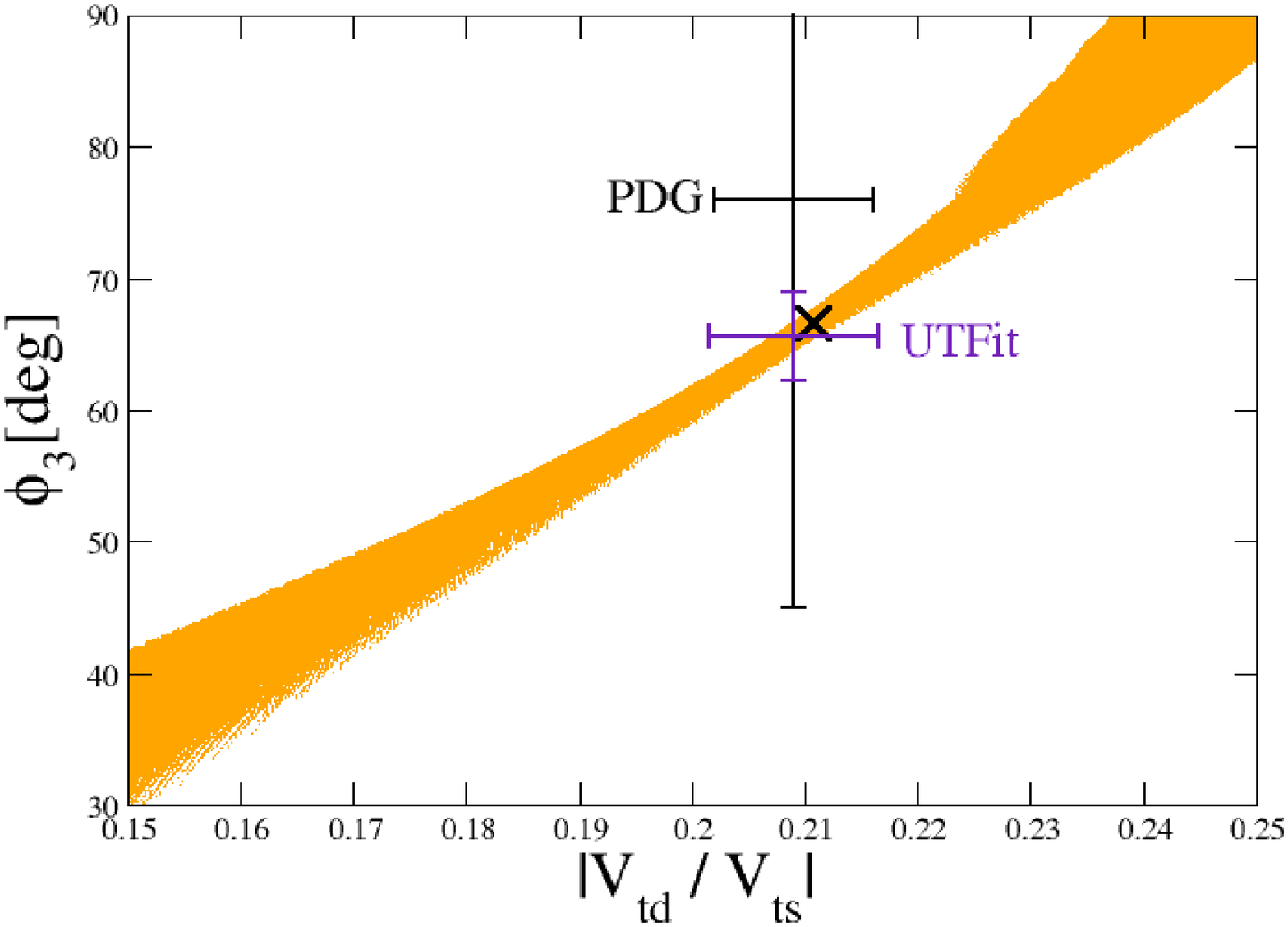}
\caption{\label{rho-eta-4}\footnotesize
The theoretical values of the model IV at $90 \%$ CL in the 
 $\bar{\rho}{\bf -}\bar{\eta}$ plane (left) and
 $|V_{td}/V_{ts}|{\bf -}\phi_3(\gamma)$ 
 plane (right).
 Two experimental values (\ref{exp-v-pdg}) and 
(\ref{exp-v-utfit}) are also plotted.
The best fit point is denoted as $\times$.
 }
\end{figure}
\begin{figure}[htb]
\includegraphics*[width=0.4\textwidth]{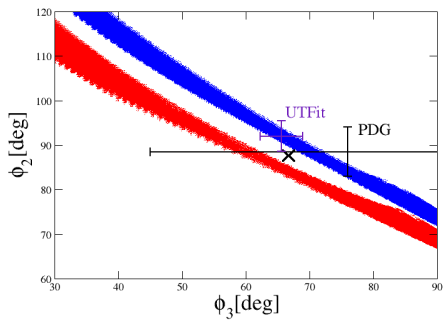}
\caption{\label{gamma-alpha-4}\footnotesize
The  theoretical values of the model IV at $90 \%$ CL in the 
 $\phi_3(\gamma){\bf -}\phi_2(\alpha)$ plane, where
two predicted regions
for $m_s/m_b=   0.0220 \pm 0.0007$ and
$m_d/m_b=( 1.200\pm 0.028) 10^{-3}$ (red) and 
for $m_s/m_b=0.0170 \pm 0.0007$ and 
$m_d/m_b=  (0.930 \pm 0.028)10^{-3}$ (blue)
are plotted.
Two experimental values (\ref{exp-v-pdg}) and 
(\ref{exp-v-utfit}) are also plotted.
The best fit point is denoted as $\times$.
}
\end{figure}

According to \cite{Harayama:1996jr}, the CKM matrix 
can be approximately written in a closed form.
One finds, for instance,
\begin{eqnarray}
  V_{us} &\simeq & -y_d \sqrt{\frac{m_d}{m_s}}
            +y_u\sqrt{\frac{m_u}{m_c}}\ e^{2i\theta}\ ,\\
 V_{cb} &\simeq & \frac{y_d^2}{\sqrt{1-y_d^4}} \frac{m_s}{m_b}\ e^{2i\theta}
            -\frac{y_u^2}{\sqrt{1-y_u^4}}\frac{m_c}{m_t}\ e^{i  \theta},\\
 V_{ub} &\simeq & \frac{\sqrt{1-y_d^4}}{y_d}
                 \sqrt{\frac{m_d}{m_s}}\frac{m_s}{m_b}\nonumber\\
              & & + y_u \sqrt{\frac{m_u}{m_c}}\left(
                    \frac{y_d^2}{\sqrt{1-y_d^4}} \frac{m_s}{m_b}\ e^{2i\theta}
                    - \frac{1}{y_u^2\sqrt{1-y_u^4}}\frac{m_c}{m_t}\ e^{i\theta}
              \right)\ ,
\end{eqnarray}
where $y_u=1/c_u$ and $y_d=1/c_d$.
For $y_d \simeq 1$ we can reproduce the classic relations of
\cite{Gatto:1968ss,Cabibbo:1968vn,Fritzsch:1977za} 
   \begin{eqnarray}
  |V_{us}| &\simeq & \sqrt{\frac{m_d}{m_s}}+O(m_u/m_c).
  \end{eqnarray}

We have performed numerical analyses on the CKM parameters
of the models in detail, some of which are presented in 
Figs.~\ref{s2b-gamma-III} -\ref{gamma-alpha-4},
where Fig.~\ref{s2b-gamma-III} is that of the model III,
and Figs.~\ref{rho-eta-4} and \ref{gamma-alpha-4} are those of the model IV.

The best fit  for the model III
is given by
\be
\sin2\phi_1(\beta) &=&0.664,~\phi_2(\alpha)=81.397^\circ , ~
  |V_{us}|=0.2250,~|V_{cb}|=0.0416,\nonumber\\
  m_u/m_t &=& 0.860\times10^{-5},~m_c/m_t=0.375\times10^{-2}, \\
  m_d/m_b &=&1.065\times10^{-3},~m_s/m_b=0.199\times10^{-1}\nonumber
\ee
with     $\chi^2 = 2.68$ for the set of parameters
$\sin2\phi_1(\beta),\phi_2(\alpha),m_u/m_t, m_c/m_t,
m_d/m_b,m_s/m_b,|V_{us}|,|V_{cb}|$.
As we can see from Fig.~\ref{s2b-gamma-III}, the model could be excluded;
the PDG value  in the $\bar{\rho}-\bar{\eta}$ plane \cite{Amsler:2008zz} is
 about $2\sigma$ away from the theoretical values.
A more precise experimental determination of $\phi_2(\alpha)$ will be more 
supporting the conclusion.

The model IV, on the other hand, seems to be consistent with
the  PDG values \cite{Amsler:2008zz} as well as with the Utfit group values \cite{utfit}.
In  Fig.~\ref{gamma-alpha-4} we plot two predicted regions
for $m_s/m_b=   0.0220 \pm 0.0007~,~
m_d/m_b=( 1.200\pm 0.028) 10^{-3}$ (red) and 
$m_s/m_b=0.0170 \pm 0.0007~,
~m_d/m_b=  (0.930 \pm 0.028)10^{-3}$ (blue),
which should be compared with (\ref{massratio}).
We see that  precise measurements of $\phi_2(\alpha)$
can distinguish the two regions.
We also find that the smaller the $m_s$ is, the larger is $\phi_2(\alpha)$\footnote{
While a smaller $m_s(M_Z)$  has been 
recently reported in \cite{Xing:2007fb},
the lattice calculation  of \cite{Blum:2007cy}
including electromagnetic interactions 
and non-perturbative renormalization effects
leads to a slightly larger $m_s(2\mbox{GeV})$
than that of \cite{Amsler:2008zz}.}.

The best fit for the set of parameters of the model IV,\\
$\bar{\rho},\bar{\eta},m_u/m_t, m_c/m_t,
m_d/m_b,m_s/m_b,|V_{us}|,|V_{cb}|$, is given by:
\be
\bar{\rho}&=&0.144,~\bar{\eta}=0.352,~
  |V_{us}|=0.2254,~|V_{cb}|=0.0411,\nn\\
 m_u/m_t &=&0.704\times10^{-5},~m_c/m_t=0.390\times10^{-2}, \\
  m_d/m_b &=&0.965\times10^{-3},~m_s/m_b=0.177\times10^{-1}\nn
\ee
with     $\chi^2 = 1.03$, and
  \be
  \phi_3(\gamma) &=&66.698^\circ,~\phi_2(\alpha)=87.650^\circ, ~
   |V_{us}|=0.2254,~|V_{cb}|=0.0411,\nn\\
  m_u/m_t &=&0.788\times10^{-5},~m_c/m_t=0.390\times10^{-2}, \\
  m_d/m_b &=&1.075\times10^{-3},~m_s/m_b=0.194\times10^{-1}\nn
\ee
with     $\chi^2 = 0.44$ for the set of parameters
$\phi_3(\gamma),\phi_2(\alpha),m_u/m_t, m_c/m_t,
m_d/m_b,m_s/m_b,|V_{us}|,|V_{cb}|$.

\section{Conclusion}
We have investigated four predictive flavor models that are based on a
low-energy discrete family symmetry. Two of them may be excluded by the present
precision of the measurements of the CKM parameters.
The problem of the model I has been already
 pointed out in \cite{Lavoura:2005kx}.
For the model II a more precise determination of the quark masses
is very crucial, especially the mass ratio $m_s/m_d$. We have found
that  the smaller 
 the mass ratio is, the better is the chance for the model II.
 We have also found that to confirm or exclude the model IV,
 more precise determinations of $\phi_3(\gamma)$ and $\phi_2(\alpha)$ as well as
 of the quark masses are indispensable. 
The uncertainty in the strange quark mass
should be less than a little more than $10\%$
to make it comparable with the assumed uncertainties
of $\sim 1^{\circ}$ and $\sim 2^{\circ}$ 
in $\phi_2(\alpha)$ and $\phi_3(\gamma)$, respectively,  at about 50 inverse atto barn
achieved at a future 
B factory.

 In this letter we have restricted ourselves to low-energy
 predictive models. There are also high-energy 
 predictive models \cite{Ma:2005tr}-\cite{Ishimori:2008fi}, which will be the  target of 
 our future investigations.

 \vspace{0.5cm}
\noindent
{\large \bf Acknowledgments}\\
We would like to thank Masashi Hazumi for instructive discussions and 
 useful comments.
This work is supported by the Grants-in-Aid for Scientific Research 
from the Japan Society for the Promotion of Science
(\# 18540257).

\end{document}